\documentstyle[preprint,aps,epsf]{revtex}
\tightenlines
\begin{document}
\draft
\title{Coulomb effects in dynamics of polar lattices}
\author{L. A. Falkovsky \footnote{e-mail: falk@itp.ac.ru}}
\address{Landau Institute for Theoretical Physics, 117337 Moscow, Russia}
\maketitle

\begin{abstract}
Zone-center phonon frequencies of polar lattices are calculated
for uniaxial crystals from the symmetry arguments.
Long-range Coulomb forces and crystal anisotropy are explicitly
taken into account.
Free-carrier contributions into a dielectric constant
are included.
The angular dispersion of optical-phonon modes
is compared with data for hexagonal 6H-SiC politype.

\end{abstract}


\section {Introduction}
Electrostatic dipole--dipole interactions play an important role in
the theory of lattice vibrations.
It is common knowledge \cite{BK}  that the degeneracy of phonon modes
at the Brillouin zone-center (e.g., in the cubic 3C-SiC crystal)
is removed if the atomic displacements are
accompanied by the Coulomb field. Then the frequency of the longitudinal
optical mode becomes larger than the frequencies of  transverse modes.
For noncubic crystals (e.g., for the hexagonal or
rhombohedral SiC polytypes),
the long-range Coulomb field gives rise also to an angular dependence of
the zone-center modes:
at ${\bf k}=0$ {\it the optical-phonon frequencies depend on the propagation
direction}.

Such a phenomenon is  rather unusual from both the physical and
 mathematical point
of view: the eigenvalues of dynamical matrix calculated for ${\bf k}=0$
depend on the ${\bf k}$-direction. This is caused by
the nonanalytic {\bf k}-dependence of the dynamical matrix which
results from the long-range dipole--dipole interaction. In polar cubic
crystals, the Coulomb field splits the three-fold degeneracy of optical modes
at the Brillouin zone-center,
but the frequency dependence on the propagation direction as well appears in
uniaxial crystals
due to the long-range electrostatic field.

The electrodynamic part of the problem was formulated by Loudon
\cite{Lou}. The Coulomb contributions in the dynamical matrix are
usually calculated be means of an Evald summation \cite{BK}.
The angular dispersion of optical modes is clearly demonstrated
by the recent
numerical calculations  of the zone-center phonons \cite{GSB} and
for the whole Brillouin zone \cite{BRB} in the case of semiconductors
 $A^3B^5$ with the wurtzite structure.
The Coulomb field is also taken  into account  in the theory of
phonon--plasmon coupled modes (polaritons) \cite{HNU}, when the effect
of free carriers is studied.

The main aim of this paper is i) to calculate the angular dispersion for
the zone-center phonons in uniaxial crystals
using the symmetry arguments and ii) to consider the effect of free
carriers on these modes. For the sake of definiteness, we are interested in
the phonon modes of uniaxial SiC polytypes  which are  very popular now in
 technical applications.

\section{optical modes at the zone-center of cubic crystals}

Among the hexagonal and rhombohedral SiC polytypes, there is the cubic
3C-SiC one with two atoms in the unit cell. First we consider the optical modes
in this simplest case. For the nearest vicinity of Brillouin
zone-center $k\ll \pi/d$, where $d$ is the lattice parameter, the acoustic
and optical modes can be divided using the series expansion in ${\bf k}$
of the dynamical matrix.
As the result, we obtain
in the zero approximation in ${\bf k}$
for the optical
displacements  $ u_i$ ($i=x,y,z$) the system of three equations
\begin{equation}  \label{op}
(\phi-M^{*}\omega^2){\bf u}={\bf f},
\end{equation}
where $M^{*}$ is the reduced mass of two atoms (Si and C) in the unit cell,
$\phi$ is the diagonal element of the force-constant matrix
(the only one diagonal element of the
$3\times 3$-matrix exist in a
cubic crystal). The value of $\phi$ can be calculated in the
nearest-neighbor approximation. As for the long-range Coulomb
interaction, it can not be considered in such a way. The Coulomb
effect is described by a force
${\bf f} = Ze {\bf E}$ acting on an effective charge
$Z$, where the electric field  ${\bf E}$
is found from  Maxwell's equations.
Eliminating the magnetic field from Maxwell's equations, one can express
the electric field ${\bf E}$ in  terms of polarization ${\bf P}$:
\begin{equation}  \label{pol}
{\bf E}=\frac{-4\pi[{\bf k}({\bf k}{\bf P})-\omega^2{\bf P}/c^2]}
{k^2-\omega^2/c^2}.
\end{equation}

We are interested in $\omega$
of the order of optical mode frequencies, i.e., $\omega/c\simeq$
10$^3$ cm$^{-1}$. If the phonon is
exited by light, its wave vector has the value of the photon wave vector,
i.e., of the order of 10$^5$ cm$^{-1}$.
Then the condition $k\gg\omega/c$ is fulfilled,
  and the terms with $c^2$ have to be omitted in Eq. (\ref{pol}) which
becomes:
\begin{equation} \label{pol1}
{\bf E}=-4\pi{\bf k}({\bf k}{\bf P})/k^2.
\end{equation}

In the long-wave limit $(k\ll \pi/d)$, the polarization is related
to the phonon displacement and electric field by the macroscopic equation:
\begin{equation} \label{pol2}
{\bf P}=NZe{\bf u} + \chi {\bf E}
\end{equation}
where $\chi$ is the atomic permittivity and
$N$ is the number of unit cells in 1 cm$^{3}$. Sometimes, the
 local field is used in equations like (\ref{pol2}) instead of
the macroscopic field ${\bf E}$. But for  cubic crystals (when the simple
 Lorentz relationship exists only),  the local field can be eliminated
renorming the force constant $\phi$.

Using Eqs. (\ref{pol1}) and (\ref{pol2}), the electric field $\bf E$
can be expressed in terms of $\bf u$. Then Eq. (\ref {op}) gives
 the frequencies
of transverse and longitudinal optical modes
in the cubic crystal
\begin{equation} \label{pol3}
\omega^2_{TO}=\phi/M^{*} \quad \text{and} \quad \omega^2_{LO}=\phi/M^{*}+\rho
\end{equation}
 where
\begin{equation} \label{def}
\rho=Z^2e^2N/\varepsilon^{\infty}\quad \text{and} \quad
 \varepsilon^{\infty}=1+4\pi\chi.
\end{equation}

Although the relation (\ref{pol1})
between ${\bf E}$ and ${\bf P}$ involves
the {\bf k}-direction explicitly,
the frequencies of optical modes (\ref{pol3})
 are independent of the propagation direction
as it must be for a cubic crystal.

\section{optical modes at the zone-center of uniaxial crystals}

The crystal anisotropy of the noncubic SiC polytypes  is known to be small
because the nearest neighbors of any given atom conserve the cubic symmetry.
Let us introduce the strain tensor $e_{ij}$
which describes a small difference between the
dynamic matrices for the noncubic polytype and the cubic one.
Then the phonon spectrum of the noncubic polytype can be obtained in the
following way. In the first step, we transform the Brillouin zone of the
cubic polytype ("the large zone") using the strain $e_{ij}$.
Hence, we find the
frequencies of the so-called strong modes. For the zone-center,
they can be obtained by the  expansion of the dynamic matrix in
the strain $e_{ij}$.

In the second step, we  take into account that noncubic polytypes
have more than two atoms in the unit cell and the additional optic modes
appear. Phonon branches of the large zone are folded \cite{Pat} into the
Brillouin zone of the noncubic polytype giving  additional weak modes.
The weak-mode intensity  in optics and Raman scattering
was calculated in \cite{Fal}.
Therefore in the present paper, we consider only  strong modes.

The dynamic matrix can contain only  components
$e_{ij}$ invariant under the symmetry transformations of the crystal.
There are  two invariants of  first order,
$e_{zz}$ and $e_{xx}+e_{yy}$,
the $z$-axis being parallel to the $c$-axis.
We can fix the crystal volume, i.e., impose the condition
$e_{ii}=0$. Then we have only one invariant, for instance,
$e_{zz}$ which can be  written only on the diagonal of the force-constant
matrix of Eq. (\ref{op}). The coefficients of the
$xx$ and $yy$ elements are equal because of the rotation invariance
 around the c-axis. At last, we can omit the common frequency shift.
Therefore instead of Eq. (\ref{op}), we obtain
\begin{equation}
\left(
\begin{array}{ccc}
\beta +\rho n_x^2-\omega ^2 & \rho n_xn_y & \rho n_xn_z \\
\rho n_xn_y & \beta +\rho n_y^2-\omega ^2 & \rho n_yn_z \\
\rho n_xn_z & \rho n_yn_z & \alpha +\rho n_z^2-\omega ^2
\end{array}
\right) \cdot \left(
\begin{array}{c}
u_x \\
u_y \\
u_z
\end{array}
\right) =0,  \label{tmac1}
\end{equation}
where ${\bf n}={\bf k}/k$ and
\begin{equation} \label{def2}
\alpha=\phi/M^{*}, \qquad
\beta =\alpha +be_{zz} .
\end{equation}
We take the vector ${\bf k}$ in the $yz$-plane and denote
$\theta$ the angle between ${\bf k}$ and the $c$-axis:
$n_x=0,n_z=\cos{\theta} $
and $n_y=\sin\theta $. Then we see from Eq. (\ref{tmac1}) that
there are  one transverse mode (TO$_1$) vibrating in $x$-direction
and  two modes in the $yz$-plane with the frequencies
\begin{eqnarray} \label{heg}
\omega _{TO1}^2=\beta ,\\
\omega _{y,z}^2(\theta )=\frac 12(\rho +\alpha +\beta )\pm\frac12
\left\{
[\rho +(\alpha-\beta)\cos{2\theta} ]^2+(\alpha-\beta
)^2\sin^2{ 2\theta} \right\} ^{1/2}.
      \nonumber
\end{eqnarray}

Emphasize that Eq. (\ref{heg}) gives the phonon frequencies in
the zone-center, but they depend on the propagation direction $\theta$.
This dependence has its origin in the simultaneous effect of the Coulomb
field (described by the constant $\rho$) and
crystal anisotropy ($\beta \ne \alpha$). In  absence of the Coulomb field
($\rho=0$), we have
$\omega _z^2=\alpha $, $\omega _y^2=\beta $, and there is no angular
dispersion. For the isotropic case
($\alpha=\beta$) ,
Eq. (\ref{heg}) gives the modes for the cubic crystal.

If the Coulomb effect is small in comparison with the crystal anisotropy
($\rho\ll|\alpha -\beta | $),
we can omit the off-diagonal terms in the matrix
(\ref{tmac1}). So there are one mode vibrating close to  the $c$-direction
with the frequency
$\omega_z^2=\alpha+\rho\cos^2\theta$ (with an accuracy to
$\rho^2/(\alpha -\beta)^2 $),
and the other mode
near the $y$-direction
with the frequency
$\omega_y^2=\beta+\rho\sin^2\theta$.

In the opposite limiting case of the small crystal anisotropy,
it is useful to pass to the coordinate system with the $z'$-axis along
the  ${\bf k}$-vector, making in Eq. (\ref{tmac1}) the unitary transformation
\begin{equation}\label{ham}
U_{ij}=\left(
\begin{array}{ccc}
1 & 0 & 0 \\
0 & \cos\theta  & \sin\theta  \\
0 & -\sin\theta  & \cos\theta
\end{array}
\right)
\end{equation}
Then we arrive to the diagonalization problem of the matrix
\begin{equation} \label{tmac2}
\left(
\begin{array}{ccc}
\beta  & 0 & 0 \\
0 & \beta \cos^2\theta +\alpha \sin^2\theta  & (\beta -\alpha )\sin\theta
\cos\theta  \\
0 & (\beta -\alpha )\sin\theta \cos\theta  & \beta \sin^2\theta +\alpha
\cos^2\theta +\rho
\end{array}
\right)
\end{equation}
We see that besides the TO$_1$ mode, in the case
$|\alpha -\beta |\ll \rho $, there are another nearly transverse  TO$_2$ mode
and the nearly longitudinal LO mode
with the frequencies
\begin{eqnarray} \label{lim}
\omega _{TO2}^2(\theta ) &=&\beta \cos^2\theta +\alpha \sin^2\theta ,
 \\ \nonumber
\omega _{LO}^2(\theta ) &=&\rho +\beta \sin^2\theta +\alpha \cos^2\theta ,
\end{eqnarray}
which can be obtained also by the expansion of Eq. (\ref{heg}) with an accuracy
to $(\alpha-\beta)^2/\rho^2$.
The dispersion curves (\ref{heg}), (\ref{lim}) are shown schematically
in Fig. 1. The angular dispersions  of the form  (\ref{lim})
were obtained by Loudon \cite{Lou}.

One can see from Eq. (\ref{heg}) that a conservation law exists. Namely,
the sum of the squared frequencies of the
$yz$ modes is independent of the propagation direction, e.g.,
\begin{equation} \label{cl}
\omega _{y}^2(\theta =0)+\omega _{z}^2(\theta =0)=\omega _{y}^2(\theta
=\pi /2)+\omega _{z}^2(\theta =\pi /2).
\end{equation}
As an example let us consider 6H-SiC polytipe. The angular dispersion of
its optical modes is known from the experiment \cite{HNU}, \cite{Pat}.
For $\theta =0$ (propagation parallel to the $c$-axis)
the TO$_1$  and  $y$ mode are degenerate, and their frequencies are equal
to $\sqrt{\beta }$. The experimental value is
 797 cm$^{-1}$ (with uncertainty about 1 cm$^{-1}$).
The corresponding value for the longitudinal mode is
$\omega _{LO}(\theta =0)=\sqrt{\rho +\alpha }$.
For $\theta =\pi /2$  (propagation perpendicular to the $c$-axis)
$\omega _{TO2}(\theta=\pi /2)=\sqrt{\alpha }$
(experimental value is 788 cm$^{-1}$) and
$\omega _{LO}(\theta =\pi /2)=\sqrt{\rho +\beta }$
(970 cm$^{-1}$). It immediately follows that: $\rho
=552.9^2$ cm$^{-2},\alpha =788^2$ cm$^{-2}$ and $\beta $=797$^2$ cm$^{-2}$.
Thus one calculates
$\omega _{LO}(\theta =0)=\sqrt{\rho +\alpha }=$962.6 cm$^{-1}$,
which should be compared with the experimental value 964 cm$^{-1}$.
The small difference
between these two values can be attributed to the anisotropy in the atomic
permittivity which is considered in the following section.

\section{effects of the permittivity anisotropy and free carriers}

We assumed in the previous section that the uniaxial anisotropy affects only
the short-range contribution to  the force-constant matrix. But in
uniaxial crystals the atomic permittivity $\chi$ is a tensor
with two independent components
$\chi_{\parallel}$  and $\chi_{\perp}$ corresponding to the crystal axes.
This effect is small because each atom has  nearly cubic
surroundings, but it should be included for a careful comparison with
experiments. In a similar way,  free carriers make a contribution
to the angular dispersion of the longitudinal optical mode.

To take into account both the
anisotropy of atomic permittivity  and the conductivity of free
carriers $\sigma$,  we write instead of Eq. (\ref{pol2}) the following
\begin{eqnarray} \label{pola}                                \nonumber
P_{\parallel}=NZe u_{\parallel} + \left(\chi_{\parallel}+i\frac{\sigma_{\parallel}}
{\omega}\right)  E_{\parallel} ,\\
P_{\perp}=NZe u_{\perp} + \left(\chi_{\perp}+i\frac{\sigma_{\perp}}{\omega}\right)
E_{\perp}.
\end{eqnarray}

Using Eqs. (\ref{pol1}) and (\ref{pola}), we obtain the equation of
motion in the form (\ref{tmac1}) and the phonon frequencies (\ref{heg}), but
the conservation law (\ref{cl}) does not work now because
$\rho$ becomes the function
of $\theta$:
\begin{equation} \label{ro}
\rho(\theta)=Z^2e^2N\left[
  \left(\varepsilon_{\parallel}^{\infty}+
4\pi i\frac{\sigma_{\parallel}}{\omega}\right)\cos^2\theta+
\left(\varepsilon_{\perp}^{\infty}+
4\pi i\frac{\sigma_{\perp}}{\omega}\right)\sin^2\theta \right]^{-1},
\end{equation}
where
$\varepsilon_{\parallel}^{\infty}=1+4\pi \chi_{\parallel},~
  \varepsilon_{\perp}^{\infty}=1+4\pi \chi_{\perp}$.
Notice, that the vibration modes obtain same
damping due to  conductivity.
In addition, the optical phonon  has a natural width $\Gamma$
given by its probability to decay into phonons of lower energy, and
the term $i\Gamma$ should be
added to $\omega$ in Eq. (\ref{tmac1}).

Then we can use  transformation  (\ref{ham}) and obtain matrix
(\ref{tmac2}) with the function $\rho(\theta)$ instead of constant $\rho$.
We see that in the limiting case of the weak anisotropy,
$|\alpha -\beta | \ll\rho(\theta)$, the Coulomb field (and therefore
the carriers) affects only
the longitudinal mode. Its frequency is determined by the equation
\begin{equation} \label{lm}
R(\omega)\equiv
\rho(\theta)+\beta \sin^2\theta +\alpha\cos^2\theta -i\omega\Gamma-\omega^2=0 ,
\end{equation}
where $\rho(\theta)$ given by Eq. (\ref{ro}) depends  on
$\omega$ explicitly and by means of the conductivity $\sigma$.

Equation (\ref{lm}) gives the frequency of the LO-phonon--plasmon coupled mode
in uniaxial semiconductors. Notice that in the isotropic case,
 Eq. (\ref{lm}) coincides with the condition
$\varepsilon (\omega)=0$, where the dielectric function
$\varepsilon(\omega)$ is given by the well-known expression
$$\varepsilon(\omega)=
\varepsilon^{\infty}\left [1+\frac{\omega_{LO}^2-\omega_{TO}^2}
{\omega_{TO}^2-\omega^2-i\omega\Gamma}
-\frac{\omega_p^2}{\omega(\omega+i\gamma)}\right],$$
and  the plasmon frequency
$$\omega_p^2=4\pi ne^2/\varepsilon^{\infty}m.$$
In this case, Eqs. (\ref{pol3}), (\ref{def}), and
(\ref{def2}) give
$\omega_{TO}^2=\alpha=\beta$,
$\omega_{LO}^2=\omega_{TO}^2+Z^2e^2N/\varepsilon^{\infty}$,
 and  the Drude formula for  conductivity reads
$\sigma= ne^2/m(-i\omega+\gamma)$.

The function $R(\omega)$ from Eq.
(\ref{lm}) is measured in  Raman experiments. Namely, the Raman intensity
as a function of frequency transfer $\omega$ is
\begin{equation} \label{ri}
 I(\omega,\theta)\simeq \text{Im}\frac{1}{R(\omega)}
\end{equation}
for   the LO mode excitation with the propagation direction $\theta$.
If the incident or scattered light has a finite aperture, Eq. (\ref{ri})
should be integrated over allowed $\theta$.

Eq. (\ref{ri}) can be used in experimental studying
the effect of carriers on the Raman
scattering in uniaxial semiconductors. Then the conductivity tensor
in  Eq. (\ref{ro}) is given by the Drude-like formula with the
diagonal components $m_{\parallel, \perp} $ and  $\gamma_{\parallel,\perp}$,
for instance,
$\sigma_{\parallel}= ne^2/m_{\parallel}(-i\omega+\gamma_{\parallel})$.

Let us emphasize the main result of the paper: the effects of crystal
anisotropy ($\alpha\ne\beta$) and  Coulomb field $\rho(\theta)$
on the phonon dispersion
are explicitly separated as one can see in Eqs. (\ref{heg}), ({\ref{lm}).

\section{acknowledgments}
This study was initiated by discussions with
J. Camassel and P. Vicente
(GES, UM2-CNRS, France), and
the author would like to thank them.
The author acknowledges  the kind hospitality of
the Max-Planck-Institute f\"{u}r Physik Komplexer Systems (Dresden)
where this work was completed.

 \newpage
{\bf Figure caption}
Angular dispersion of optical-phonon modes in uniaxial crystals at
the zone-center. The angle $\theta$ is the angle between the c-axis and
the wave vector ${\bf k}\rightarrow 0$. The TO$_1$ mode is polarized
perpendicular to the c-${\bf k}$ plane. The LO and TO$_2$ modes have a
nearly longitudinal  and transverse character, respectively,
if the Coulomb force effects dominate over the crystal anisotropy.
\newpage
 \epsfxsize=120mm
 \epsfysize=120mm
 \centerline{\epsfbox{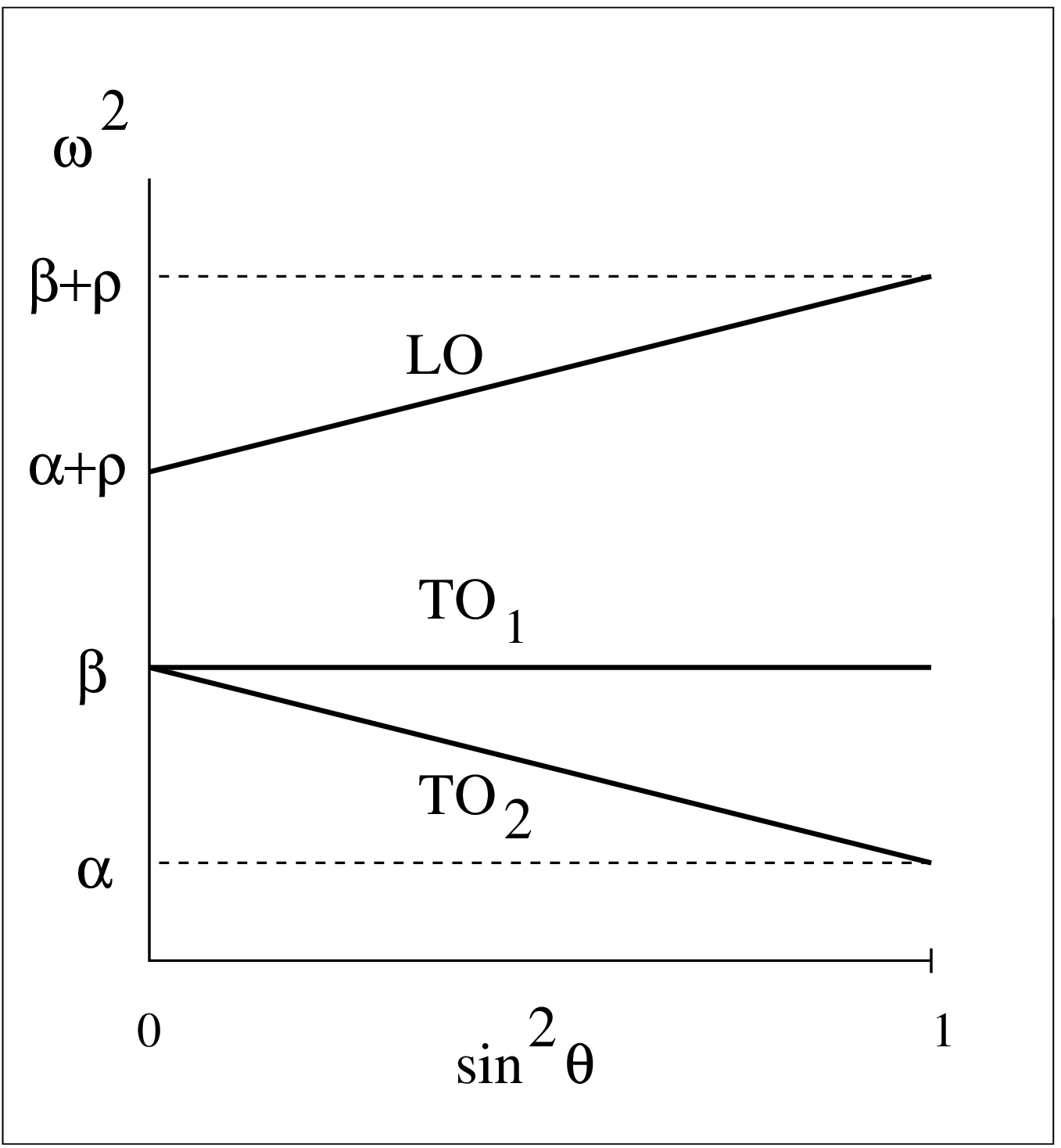}}

\end{document}